\documentstyle[12pt,a4,cite]{article}

\renewenvironment{thebibliography}[1]
{\small
 \begin{list}{[\arabic{enumi}]}
 {\usecounter{enumi} \setlength{\parsep}{0pt}
  \setlength{\itemsep}{0pt} \settowidth{\labelwidth}{[#1]}
  \settowidth{\leftmargin}{[100]}
  \sloppy}}
 {\end{list}}

\renewcommand{\theequation}{\arabic{equation}}

\def\to{\rightarrow}
\def\epem{\ifmmode{ e^{+}e^-} \else{$ e^{+}e^- $ } \fi}
\def\ttbar{\ifmmode{t\bar{t}} \else{$t\bar{t}$} \fi}

\def\simlt{\rlap{\lower 3.5 pt \hbox{$\mathchar \sim$}}%
 \raise 1pt \hbox {$<$}}
\def\simgt{\rlap{\lower 3.5 pt \hbox{$\mathchar \sim$}}%
 \raise 1pt \hbox {$>$}}
\def\sp3{\ifmmode{\:\:\:}\else{$\:\:\:$} \fi}
\def\p3{\:\:\:}
\def\3h{\:\:\:\,}
\def\4{\:\:\:\:}
\def\5{\:\:\:\:\:}
\def\6{\:\:\:\:\:\:}
\def\mm{\!\!}
\def\m5{\!\!\!\!\!}

\def\ebar{\bar{e}}
\def\sbar{\bar{s}}
\def\cbar{\bar{c}}
\def\gzbar{\bar{g}_Z}
\def\gwbar{\bar{g}_W}

\def\ehat{\hat{e}}
\def\shat{\hat{s}}
\def\chat{\hat{c}}
\def\gzhat{\hat{g}_Z}
\def\ghat{\hat{g}}

\def\pibar{\overline{\Pi}}
\def\msbar{\overline{\rm MS}}
\def\gambar{\overline{\Gamma}}
\def\delb{\bar{\delta}_{b}}
\def\delg{\bar{\delta}_{G}}

\makeatletter
\newtoks\@stequation

\def\subequations{\refstepcounter{equation}%
  \edef\@savedequation{\the\c@equation}%
  \@stequation=\expandafter{\theequation}
  \edef\@savedtheequation{\the\@stequation}
  \edef\oldtheequation{\theequation}%
  \setcounter{equation}{0}%
  \def\theequation{\oldtheequation\alph{equation}}}

\def\endsubequations{%
  \ifnum\c@equation < 2 \@warning{Only \the\c@equation\space subequation
    used in equation \@savedequation}\fi
  \setcounter{equation}{\@savedequation}%
  \@stequation=\expandafter{\@savedtheequation}%
  \edef\theequation{\the\@stequation}%
  \global\@ignoretrue}

\def\eqnarray{\stepcounter{equation}\let\@currentlabel\theequation
\global\@eqnswtrue\m@th
\global\@eqcnt\z@\tabskip\@centering\let\\\@eqncr
$$\halign to\displaywidth\bgroup\@eqnsel\hskip\@centering
     $\displaystyle\tabskip\z@{##}$&\global\@eqcnt\@ne
      \hfil$\;{##}\;$\hfil
     &\global\@eqcnt\tw@ $\displaystyle\tabskip\z@{##}$\hfil
   \tabskip\@centering&\llap{##}\tabskip\z@\cr}

\makeatother

\newcommand{\pr}{\hspace{\parindent}}

\newcommand{\beq}{\begin{equation}}
\newcommand{\eeq}{\end{equation}}
\newcommand{\bea}{\begin{eqnarray}}
\newcommand{\eea}{\end{eqnarray}}
\newcommand{\bsub}{\begin{subequations}}
\newcommand{\esub}{\end{subequations}}

\begin{document}

\pagestyle{plain}
\thispagestyle{empty}

\vspace*{-10mm}

\baselineskip12pt
\begin{flushright}
\begin{tabular}{l}
{\bf KEK-TH-369 }\\
{\bf SNUTP 93-50 }\\
{\bf YUMS  93-19 }\\
{\bf KEK preprint 93--108 }\\
October 1993
\end{tabular}
\end{flushright}

\baselineskip18pt
\vspace{8mm}
\begin{center}
{\Large \bf Electroweak Radiative Corrections } \\
\vglue 8mm
{\bf K.~HAGIWARA,~~~S.~MATSUMOTO } \\
\vglue 1mm
{\it Theory Group, KEK, Tsukuba, Ibaraki 305, Japan }\\
\vglue 4mm
{and} \\
\vglue 4mm
{\bf C.S.~KIM }\\
\vglue 1mm
{\it Department of Physics, Yonsei University, Seoul 120-749, Korea }\\
\vglue 20mm
{\bf ABSTRACT} \\
\vglue 10mm
\begin{minipage}{14cm}
{\normalsize
A new framework to study electroweak physics at one-loop level
in general ${\rm SU(2)_L \times U(1)_Y}$ theories is introduced.
It separates the 1-loop corrections into two pieces:
process specific ones from vertex and box contributions and the
universal ones due to contributions to the gauge boson propagators.
The latter are parametrized in terms of four effective form factors,
$\bar{e}^2(q^2)$, $\bar{s}^2(q^2)$, $\bar{g}_Z^2(q^2)$,
and $\bar{g}_W^2(q^2)$, correspondingly to $\gamma\gamma$,
$\gamma Z$, $ZZ$, and $WW$ propagators.
By assuming only the standard model contributions to the process
specific corrections, the magnitudes of the four form factors are
determined at $q^2=0$ and at $q^2=m_Z^2$ from all available
precision experiments.
These values are then compared systematically with the predictions
of the ${\rm SU(2)_L \times U(1)_Y}$ theories.
No deviation from the standard model has been identified.
Plausible range of the top quark mass is then obtained
for a given Higgs boson mass and $\alpha_s$ .
}
\end{minipage}
\end{center}
\vglue 20mm
\noindent
\baselineskip14pt
--------------------------------------------------------------------------\\
Plenary talk presented by K. Hagiwara at the {\it 14'th International Workshop
on Weak Interactions and Neutrinos}, July 19-24, 1993,
Seoul National University, Seuol, Korea.
To be published in the proceedings.
\vfill
\newpage
\setcounter{page}{1}
\begin{center}
{\bf ELECTROWEAK RADIATIVE CORRECTIONS } \\
\vglue 8mm
{\bf K.~HAGIWARA,~~~S.~MATSUMOTO } \\
\vglue 1mm
{\it Theory Group, KEK, Tsukuba, Ibaraki 305, Japan }\\
\vglue 4mm
{and} \\
\vglue 4mm
{\bf C.S.~KIM }\\
\vglue 1mm
{\it Department of Physics, Yonsei University, Seoul 120-749, Korea }\\
\vglue 8mm
{\bf ABSTRACT} \\
\vglue 4mm
\baselineskip4pt
\begin{minipage}{14cm}
{\small
A new framework to study electroweak physics at one-loop level
in general ${\rm SU(2)_L \times U(1)_Y}$ theories is introduced.
It separates the 1-loop corrections into two pieces:
process specific ones from vertex and box contributions and the
universal ones due to contributions to the gauge boson propagators.
The latter are parametrized in terms of four effective form factors,
$\bar{e}^2(q^2)$, $\bar{s}^2(q^2)$, $\bar{g}_Z^2(q^2)$,
and $\bar{g}_W^2(q^2)$, correspondingly to $\gamma\gamma$,
$\gamma Z$, $ZZ$, and $WW$ propagators.
By assuming only the standard model contributions to the process
specific corrections, the magnitudes of the four form factors are
determined at $q^2=0$ and at $q^2=m_Z^2$ from all available
precision experiments.
These values are then compared systematically with the predictions
of the ${\rm SU(2)_L \times U(1)_Y}$ theories.
No deviation from the standard model has been identified.
Plausible range of the top quark mass is then obtained
for a given Higgs boson mass and $\alpha_s$ .
}
\end{minipage}
\end{center}
\vglue 10mm
\baselineskip14pt

\section*{\normalsize {\bf 1.
PRECISION EXPERIMENTS CONFRONT SUSY-GUT }}

\pr
One of the most exciting developments of recent years has been the
observation~\cite{sgut91} that the electroweak mixing angle
$\sin^2\theta_W$ measured precisely at LEP agrees excellently with
the prediction of the supersymmetric (SUSY) SU(5) grand unification
theory (GUT).
The agreement is so impressive that we can hope in the near future
to learn about SUSY particle masses\cite{hmy,hy1} with a better
measurement of the QCD coupling constant.
It has been argued \cite{bh} that the uncertainty in the GUT
scale particle masses screens any possible effects of SUSY particle
threshold corrections to the coupling constant unification condition.
The works of refs.~\cite{hmy,hy1} showed that the non-observation of
proton decay effectively constrains the GUT particle contributions
to the coupling constant unification and that our hope of learning
about the SUSY mass scale from the precision measurements has been revived.


In fact we can tell more about these new particles,
about their quantum numbers, if this grand unification of the three
couplings is not merely an accident.
This is because the old SU(5) model, the grand unification model
without new particles, predicts too small a value of
$\sin^2\theta_W$ $(\sim 0.21)$
and at the same time too rapid proton decay $(\tau_p \sim 10^{29} years)$,
which are both clearly inconsistent with experiments.
It is instructive to note that the prediction for $\sin^2\theta_W$
can change only by the introduction of an incomplete SU(5) multiplet;
the only example in the SM other than the gauge bosons themselves
is the Higgs doublet, whose triplet partner of the SU(5) 5-plet
should be super-heavy.
A simple exercise shows that we can reproduce roughly the
observed values of $\sin^2\theta_W$ $(\sim 0.23)$ by simply introducing
five more Higgs doublets to the SM.
This, however, necessarily leads to further shortening of the
proton lifetime
since the unification scale $(\sim m_X)$ decreases significantly as we
introduce more Higgs doublets.  We can intuitively understand this
trend because the Higgs bosons make the SU(2) and U(1) couplings
larger at high energies, while the SU(3) coupling remains unaffected
by them.  The point at which the three couplings meet should hence
be lower in the energy scale.  In order to raise the unification
point to avoid rapid proton decay, we should also make the SU(3)
coupling large at high energies.  We therefore need new colored particles
as well at the electroweak scale,
such as scalar gluons or leptoquarks\cite{split}.
It is instructive to note that the six Higgs doublets as found
above are exactly what the minimal SUSY-SM effectively predicts,
with its two Higgs doublets each accompanied by fermionic partners,
because each fermionic partner of the scalar doublet contributes
twice as much to the running of the gauge coupling constants.
Furthermore, the colored particles that are necessary to enlarge
the proton lifetime are supplied in this model as gluinos and squarks.
We observe here that the minimal SUSY-SM, and hence the SUSY-SU(5) model,
very naturally satisfies the two experimental requirements of the
coupling constant unification.

What is exciting about this exercise is that there now seems to
be a strong indication that new particles and new interactions
among them may exist at the electroweak scale.
They may be produced at the Tevatron, LEP2 and at super colliders.
Their effects could be observed in precision experiments through quantum
corrections prior to their discovery.
The effects can be significant if some of the new particles are
as light as weak bosons or if there exist new strong interactions
among them.
Even in the absence of such a signal, we will learn more about
possible new particle spectra from the future precision
experiments.
This is our motivation to study electroweak radiative corrections
and we developed recently a new approach that allows us to look for
new physics effects systematically~\cite{hhkm}.
We hope that the crucial roles of the future precision experiments
at both high and low energies are made clear in this new framework.

\section*{\normalsize {\bf 2.
A NEW FRAMEWORK FOR 1-LOOP ELECTROWEAK PHYSICS }}

\pr
Since what we want to learn from the electroweak precision
experiments are the possible effects of new physics beyond the SM,
whose exact nature is unknown, we would like to analyse the data
in a framework which allows interpretations in wider classes of
theoretical models.
On the other hand the framework cannot be too general,
since our ability to identifying effects of new physics from
the precision experiments relies on the renormalizability of the
electroweak theory which allows us to predict many observables
in terms of a few parameters up to finite quantum corrections.
Because the SM corrections are precisely known,
those experiments which are sensitive to the quantum effects have
a chance to identify a signal of physics beyond the SM.
We therefore restrict ourselves to models that respect
${\rm SU(2)_L\times U(1)_Y}$ gauge symmetry which breaks
spontaneously down to ${\rm U(1)_{EM}}$.
In our approach, all new physics contributions that do not respect
the spontaneously broken ${\rm SU(2)_L \times U(1)_Y}$ gauge symmetry
can be identified by our inability to fit the data successfully
within our framework:  these exotic interactions include
all non-renormalizable effective
interactions among light quarks and leptons that may arise from
an exchange of a heavy particle such as a new gauge boson or
leptoquark boson, or from new strong interactions that bind
common constituents of quarks and leptons.

Our restriction
on the electroweak gauge group implies
in the tree level that all quarks and
leptons couple to the electroweak gauge bosons universally with
the same coupling constant as long as they have common electroweak
quantum numbers.
This universality of the gauge boson coupling to quarks and leptons
can in general be violated at the quantum level.
It has widely been recognized, however, that this universality of
the couplings holds true even in the one-loop level in a wider
class of models where new particles affect the precision
experiments only via their effects on the electroweak gauge
boson propagators\cite{lps,kl,kms,lep1,stu,stu_mod,ab,gw}.
This class of effects due to new physics is often called
oblique\cite{lps,stu} or propagator\cite{ab} corrections or
those satisfying generalized universality\cite{gw}.
This concept of universality can be generalized to certain vertex
corrections with the non-standard weak boson interactions\cite{hisz}.
It is also often useful in theories with intrinsic vertex and box
corrections, such as the SUSY-SM, since the propagator corrections
are generally larger than the vertex/box ones:  propagator corrections
can be significant either because of a large multiplicity of contributing
particles or by a presence of a relatively light new particle, whereas
the vertex and box corrections depend on a specific combination
of new particles that match the quantum number of the process
and are suppressed if one of them is heavy.

Our framework adopts this distinction between new physics
contributions to the gauge boson propagators and the rest,
where we allow the most general contributions in the former whereas we
consider only the SM contributions to the latter
(vertex and box corrections).
The new physics degree of freedom is then expressed in terms of
four charge form factors, each associated with the four types of the
electroweak gauge boson propagators :
\bsub
\label{barcharges}
\bea
	\ebar^2(q^2) &=& \ehat^2
	[1 -{\rm Re}\, \overline{\Pi}_{T,\gamma}^{\gamma\gamma}(q^2)]
	\:\:\:\:\: \mbox{for the $\gamma\gamma$ propagator},
\\
	\sbar^2(q^2) &=& \shat^2
	[1 +{\chat \over \hat{s}}
	{\rm Re}\, \overline{\Pi}_{T,\gamma}^{\gamma Z}(q^2)]
	\:\: \mbox{for the $\gamma Z$ propagator},
\\
	\gzbar^2(q^2) &=& \gzhat^2
	[1 -{\rm Re}\, \overline{\Pi}_{T,Z}^{ZZ}(q^2)]
	\:\:\:\: \mbox{for the $ZZ$ propagator},
\\
	\gwbar^2(q^2) &=& \ghat^2
	[1 -{\rm Re}\, \overline{\Pi}_{T,W}^{WW}(q^2)]
	\:\:\: \mbox{for the $WW$ propagator}, \vphantom{\frac{\chat}{\shat}}
\eea
\esub
where the hatted couplings $\ehat=\ghat\shat=\gzhat\shat\chat$
and the propagator functions are renormalized in the $\msbar$ scheme.
In addition to these four form factors we have the two weak boson
masses $m_W$ and $m_Z$ as the parameters of the electroweak theory.
Since the charge form factors are real continuous functions of
$q^2$, we have infinite degrees of free parameters when we use
them to parametrize a theory.  In practice, however, these charge
form factors can be measured accurately enough only at specific
$q^2$ ranges;  all four of them at $q^2=0$ ($q^2 \ll m_Z^2$), and
two of them, $\sbar^2(q^2)$ and $\gzbar^2(q^2)$, at $q^2 = m_Z^2$.
Hence, we have just 8 parameters that are measured accurately to test
a theory.
Among these 8 parameters, three are known precisely;
$\alpha$, $G_F$ and $m_Z$.
Since the gauge boson properties are fixed at tree level by the
three parameters in general models with the ${\rm SU(2) \times U(1)}$
symmetry broken by a vacuum expectation value, we can use the remaining
5 parameters to test the theory at the quantum level:  see Table~1.
We therefore first determine the 5 parameters,
$\sbar^2(m_Z^2)$, $\gzbar^2(m_Z^2)$, $\sbar^2(0)$, $\gzbar^2(0)$,
and $\gwbar^2(0)$, from precision experiments, and then confront
their values with various theoretical predictions.
\begin{center}
{\footnotesize Table~1}
\end{center}
\begin{center}
\begin{tabular}{|l|l|l|l|}
  \hline
   & $\begin{array}{l}
         \mbox{accurately measured} \\
         \mbox{parameters}
      \end{array}$
   & $\begin{array}{l}
         \mbox{precisely known} \\
         \mbox{parameters}
      \end{array}$
   & $\begin{array}{l}
         \mbox{parameters} \\
         \mbox{to test a theory}
      \end{array}$
  \\
  \hline
    \begin{picture}(60,16)(0,-4)
    \end{picture}
   & \hphantom{$m_W$} $\;\bar{e}^2(0)$ $\quad\;\:*$
   & $\alpha = \bar{e}^2(0)/4\pi$
   & \hphantom{$\bar{g}_W^2(0)$} $\;\;\:\:\:*$
  \\
  \hline
    \begin{picture}(60,16)(0,-4)
    \end{picture}
   & \hphantom{$m_W$} $\;\bar{s}^2(0)$ $\;\;\bar{s}^2(m_Z^2)$
   &
   & $\bar{s}^2(0)$  $\;\;\bar{g}_Z^2(m_Z^2)$
  \\
  \hline
    \begin{picture}(60,16)(0,-4)
    \end{picture}
   & $m_Z\,$ $\;\bar{g}_Z^2(0)$  $\;\bar{g}_Z^2(m_Z^2)$
   & $m_Z$
   & $\bar{s}^2(0)$  $\;\;\bar{g}_Z^2(m_Z^2)$
  \\
  \hline
    \begin{picture}(60,16)(0,-4)
    \end{picture}
   & $m_W$  $\:\bar{g}_W^2(0)$  $\quad\:*$
   & $4\sqrt{2}G_F ={\bar{g}_W^2(0) \over m_W^2}\,(1+\bar{\delta}_G)$
   & $\bar{g}_W^2(0)$  $\;\;\:\:\:*$
  \\
  \hline
\end{tabular}
\end{center}

When the new physics scale is significantly higher than the scale
($\simlt m_Z^2$) of precision measurements, we can often neglect new
physics contributions to the running of the charge form factors.
Among our 5 parameters, the values of $\sbar^2(0)$ and $\gzbar^2(0)$
can then be determined from $\sbar^2(m_Z^2)$ and $\gzbar^2(m_Z^2)$,
respectively, by the SM physics only.
The effective number of the free parameters is then 3,
which corresponds precisely to that of
$S$, $T$, $U$\cite{stu},
$\epsilon_1$, $\epsilon_2$, $\epsilon_3$\cite{ab},
or other related triplets of parameters in refs.\cite{stu_mod}.
When the scale of new physics that couples to gauge boson
propagators is near the weak boson masses, we can identify its
signal as an anomalous running of the charge form factors.
This point has been emphasized in refs.\cite{susy_thr}
in connection with possible existence of the light SUSY particles.
The triplet parametrizations are then no longer sufficient to account
for new physics degrees of freedom, and we should regard all 5 parameters
in the Table~1 as free parameters.
Several alternative approaches to the same problem have been proposed
in refs.\cite{susy_thr,abc,stuvw}.

Even if new physics scale is large, there can appear an anomalous
running of the charge form factors.  In fact, when new physics
is parametrized in terms of the 4 gauge invariant dimension-six
operators\cite{gw}, $O_{DB}$, $O_{DW}$, $O_{BW}$, and $O_{\phi,1}$
in the notation of \cite{hisz}, then the new physics associated
with the operators $O_{DB}$ and $O_{DW}$ contribute to the running
of all the charge form factors\cite{hisz}.  Hence the operator
formalism of ref.\cite{gw} is not equivalent to the approaches with
3 oblique parameters\cite{stu,stu_mod,ab}, but is comfortably accommodated
within our framework.

A clear advantage of this approach is that we can test the
electroweak theory at qualitatively different levels.
If we find an inability to fit all the data at a given $q^2$ with
common form factor values, we should either
look for new physics that affect the relevant vertex/box corrections
significantly or else we should introduce new tree level interactions
such as those induced by an exchange of a new heavy boson.
If the 'universality' in terms of the above four charge form factors holds,
but their $q^2$-dependence does not agree with the expectations of the
standard model, we may anticipate a new physics scale very near to the
present experimental limit.
Hence new physics contributions that decouple at low energies
($\sim q^2/M^2$) can be identified as an anomalous running of
the charge form factors.
Finally, if even the running of the form factors is found to be
consistent with the SM, then our approach reduces to the standard
three parameter analyses\cite{stu,stu_mod,ab}.
Deviation from the SM is still possible since the SM has only
two free parameters, $m_t$ and $m_H$.
Here we have sensitivity to those new physics contributions that
do not decouple at low energies.

In the minimal SM, all the quantum corrections are determined
by just two parameters, $m_t$ and $m_H$, and hence all the
charge form factors are determined by their values.
We show in Fig.~1 the four charge form factors in the SM.
The trajectories are fixed such that they give correct values for
the 3 precisely known parameters, $\alpha$, $G_F$, and $m_Z$.
We show 12 trajectories each for six combination of the mass
values, $m_t=$~100, 150, 200~GeV, $m_H=$~100, 1000~GeV, and for
space-like ($q^2<0$) and time-like ($q^2>0$) momenta.  The electric
charge form factor $\bar{\alpha}(q^2)=\ebar^2(q^2)/4\pi$ does not
depend on $m_H$.  The threshold singularities are clearly seen
in the time-like trajectories.  Light hadron threshold effects do not
show up since we adopt the dispersion integral fit of the hadronic
contributions to the vacuum polarizations in the space-like
region\cite{pi3q_h,piqq_h} also for their contribution in the time-like
region.
\begin{center}
\vspace{12cm}
\begin{minipage}{14cm}
\begin{center}
{\footnotesize Fig.~1:  Charge form factors in the minimal SM. }
\end{center}
\end{minipage}
\end{center}

The 5 parameters that we determine from precision experiments are
also shown as 'data' points in the figures.  It is clear that
these 'data' are perfectly consistent with the predictions of the
minimal SM, for a certain ($m_t$, $m_H$) range, and that no indication
of new physics is found.
It should be noted here that there is no good measurement of the charge
form factors $\ebar^2(q^2)$ and $\gwbar^2(q^2)$ except at low energies
$q^2 \sim 0$.  We may expect TRISTAN and HERA to measure them and the
$W$ widths measure $\gwbar^2(m_W^2)$, but it is challenging to achieve
an accuracy comparable to those achieved in the low energy neutral current
experiments ($\sbar^2(0)$ and $\gzbar^2(0)$).

We now explain some technical details of our framework.
Those who are interested only in the results of our analysis may skip
to section 3.

The gauge boson two-point functions that appear in eq.(1) are defined
as follows:
\beq
\label{pitv}
	\pibar_{T,V}^{AB}(q^2)
	= { \pibar_T^{AB}(q^2) -\pibar_T^{AB}(m_V^2) \over q^2 -m_V^2 },
\eeq
where $m_V$ is the pole mass of the gauge boson $V$ ($m_\gamma =0$)
and the subscript $T$ stands for the transverse part of the vacuum
polarization tensor $\Pi_{\mu \nu}(q)$.  The effective charge form
factors of eq.(1) naturally appear in the $S$-matrix elements of
the gauge boson exchange processes with external light fermions,
which can be shown schematically as follows.
The Dyson summation of the one particle irreducible propagator factors
$\Pi^{VV}_T(q^2)$ gives the full $VV$ propagator at the one-loop level
\beq
\label{gtvv_b}
	G^{VV}_T(q^2)
	= \frac{1}{ q^2 -\hat{m}_V^2 +\Pi_T^{VV}(q^2) },
\eeq
where $\hat{m}_V$ is the bare mass of the vector boson $V$.
The physical mass and the width is then obtained as the pole position
of the above full propagator:
\beq
\label{polemass}
	m_V^2 -im_V\Gamma_V = \hat{m}_V^2 -\Pi_T^{VV}(m_V^2 -im_V\Gamma_V) ,
\eeq
which can be solved perturbatively.
Consistent perturbative expansion of the full propagator is then
obtained as
\beq
\label{gtvv_r}
	G^{VV}_T(q^2)
	= \frac{1}{ q^2 -m_V^2 +im_V\Gamma_V } \{1 -\Pi_{T,V}^{VV}(q^2) \},
\eeq
and the $S$-matrix elements contain the effective charge factors of eq.(1).

The propagators are calculated in the 'tHooft-Feynman gauge and
the so-called pinch term\cite{kl,pt1,pt2} of the vertex functions due to
diagrams with the weak boson self-couplings are included in the overlined
functions $\pibar_T^{AB}(q^2)$:
\bsub
\label{pibar}
\bea
	\pibar_T^{\gamma\gamma}(q^2) &=& \Pi_T^{\gamma\gamma}(q^2)
	-\frac{\ehat^2}{4\pi^2} q^2 B_0(q^2;m_W,m_W),
\\
	\pibar_T^{\gamma Z}(q^2) &=& \Pi_T^{\gamma Z}(q^2)
	-\frac{\ehat\gzhat\chat^2}{8\pi^2} (2q^2-m_Z^2) B_0(q^2;m_W,m_W),
\\
	\pibar_T^{ZZ}(q^2) &=& \Pi_T^{ZZ}(q^2)
	-\frac{\gzhat^2\chat^4}{4\pi^2} (q^2-m_Z^2) B_0(q^2;m_W,m_W),
\\
	\pibar_T^{WW}(q^2) &=& \Pi_T^{WW}(q^2)
	-\frac{\ghat^2}{4\pi^2} (q^2-m_W^2)
	[\chat^2 B_0(q^2;m_W,m_Z) +\shat^2 B_0(q^2;m_W,m_\gamma)].
\eea
\esub
Here $B_0$ is a Passarino-Veltman propagator function\cite{pv} in
the notation of ref.~\cite{dhy}.  There are two advantages in absorbing
the above $q^2$ dependent propagator-like parts of the vertex functions
into the effective charges\cite{kl}, as compared to the standard
ones\cite{lep1} that absorb the relevant vertex term at zero momentum transfer.
One is that the remaining vertex parts no more give rise to large
logarithms of the type $\ln(-q^2/m_W^2)$, and hence the effective
charges are useful in making the improved Born approximation\cite{kl}
even at very high energies ($|q^2| \gg m_W^2$).
The second is that the effective charges are now gauge invariant\cite{kl,pt2},
and hence we can discuss their properties independently of the other
corrections of the same order which are process specific.
Most importantly, we can obtain explicitly renormalization group
invariant relations between the $\msbar$ couplings
and the effective charges
\bsub
\label{rg}
\bea
	\frac{1}{\ebar^2(q^2)} &=& \frac{1}{\ehat^2(\mu)}
	[1 +Re\pibar_{T,\gamma}^{\gamma\gamma}(q^2)],
\\
	\sbar^2(q^2) &=& \shat^2(\mu)
	+\frac{\ebar^2(q^2)}{\ehat(\mu)\gzhat(\mu)}
	Re\pibar_{T,\gamma}^{\gamma Z}(q^2),
\eea
\esub
within the 'tHooft-Feynman gauge of the electroweak theory.
The trajectories of all the $\msbar$ couplings
($\ehat=\ghat\shat=\gzhat\shat\chat$) are completely fixed
by the above two equations, which can be used to study quantitatively
the heavy particle threshold corrections in GUT theories\cite{hy1,yy1}.

In our analysis we adopt the $\msbar$ couplings as the expansion parameters
of the perturbation series, since we find them most convenient when
studying consequences of various theoretical models beyond the SM.
Their usefulness in the SM analysis has been emphasized in ref.\cite{msb_ew}.
However, it is not convenient to use the $\msbar$ couplings at a specific
unit-of-mass ($\mu$) scale, such as $\mu=m_Z$, when dealing with a theory
with particles much heavier than the weak bosons because of the appearance
of large logarithms of their masses.  We hence take the following
renormalization conditions
\beq
\label{rc}
	\ehat^2 =\ebar^2(m_Z^2), \5 \shat^2 =\sbar^2(m_Z^2),
\eeq
consistently for all processes that we study.  The above conditions
renormalize all the logarithms of large masses with the help of the
renormalization group identities (\ref{rg}).
We note here that the running of $\ebar^2(q^2)$ and $\sbar^2(q^2)$
at low energies as observed in Fig.~1 is due to
the QED interactions\cite{4f_rc}, and hence the ratio
$\ebar^2(q^2)/\sbar^2(q^2)$ is not an appropriate expansion parameter
of the weak corrections even at $|q^2|\ll m_Z^2$.
We further note that, apart from details concerning the higher order
terms, our effective charges $\ebar^2(q^2)$ and $\sbar^2(q^2)$ are
the same as the real parts of the corresponding star-scheme\cite{kl}
charges, $e_*^2(q^2)$ and $s_*^2(q^2)$, respectively.

\section*{\normalsize {\bf 3.
PRECISION EXPERIMENTS }}
\pr
All the precision experiments that are sensitive to electroweak physics
at the one loop level have so far been confined to those processes with
external light quarks and leptons, where their masses can safely be
neglected as compared to the weak boson masses.
They are the $Z$ boson properties as measured at LEP and SLC, the neutral
current ($NC$) processes at low energies ($\ll m_Z$), the charged current
($CC$) processes at low energies and the $W$ mass measurements at the
$p\bar{p}$ colliders.
The relevant observables in these processes are expressed in terms of
the $S$-matrix elements of four external light fermions which form a scalar
product of two chirality conserving currents.
All the information on the electroweak physics can be learned by studying
the scalar amplitude multiplying these current-current products.

For example, we parameterize the $S$-matrix element of the $NC$ process
$ij \to ij $ (or any one of its crossed channels) as
\beq
\label{t_ij}
	T_{ij} = M_{ij} J_i \cdot J_j,
\eeq
where $J^\mu_i$ and $J^\mu_j$ denote the bare currents without the coupling
factor:
$J^\mu_i = \bar{\psi}_f\gamma^\mu P_\alpha \psi_f$ for $i=f_\alpha$, where
$P_L = (1-\gamma_5)/2$ and $P_R = (1+\gamma_5)/2$ are the chiral projectors.
All the one-loop corrections appear in the scalar amplitude $M_{ij}$ which
depends on the invariant momentum transfers $s$ and $t$.

In the neutral current amplitudes, the photonic corrections attached only
to the external fermion lines are gauge invariant in themselves.
Therefore we can obtain finite and gauge invariant amplitudes by excluding
all the external photonic corrections.
By using the charge form factors of eq.(\ref{barcharges}),
we find e.g. for the process $i\bar{i}\to j\bar{j}$
\bea
	M^{NC}_{ij} &=& \frac{Q_iQ_j}{s}
	[\ebar^2(s) +\ehat^2(\Gamma^i_1 +\Gamma^j_1)(s)
	-i\ehat^2 Im\pibar^{\gamma\gamma}_{T,\gamma}(s) ]
	+\ehat^2[(Q_iI_{3j}) \frac{\gambar_2^j(s)}{s}
	        +(I_{3i}Q_j) \frac{\gambar_2^i(s)}{s} ]\mbox{\hspace{15mm}}
\nonumber\\
	&&\; +\frac{1}{s-m_Z^2 +is\frac{\Gamma_Z}{m_Z}} \{
	(I_{3i}-Q_i\shat^2)(I_{3j}-Q_j\shat^2)
	[\gzbar^2(s) +\gzhat^2(\Gamma^i_1 +\Gamma^j_1)(s)
	-i\gzhat^2\Delta_Z(s) ]
\nonumber\\
	&&\hspace{13mm} +(I_{3i}-Q_i\shat^2)\gzhat^2
	[I_{3j}(\chat^2\gambar_2^j +\Gamma_3^j)(s) +\Gamma_4^j(s)
	-Q_j(\sbar^2(s)-\shat^2 +i\shat\chat\Delta_{\gamma Z}(s))]
\nonumber\\
	&&\hspace{13mm} +(I_{3j}-Q_j\shat^2)\gzhat^2
	[I_{3i}(\chat^2\gambar_2^i +\Gamma_3^i)(s) +\Gamma_4^i(s)
	-Q_i(\sbar^2(s)-\shat^2 +i\shat\chat\Delta_{\gamma Z}(s))] \}
\nonumber
\eea
\beq
\label{m_nc}
	+B^{NC}_{ij}(s,t). \mbox{\hspace{105mm}}
\eeq
The vertex functions $\Gamma_n^{f_\alpha}(s)$ and the box functions
$B_{f_\alpha f'_\beta}(s,t)$ are process specific.
We first note that the residues of the $\gamma$ and $Z$ poles are separately
physical observables ($\mu$-independent and gauge invariant).  At $q^2 = 0$,
we find
\beq
\label{ward}
	\Gamma_1^{f_\alpha}(0) =\gambar_2^{f_\alpha}(0) =0
\eeq
for all $f_\alpha$, which are ensured by the Abelian and non-Abelian parts
of the Ward identities, respectively.
The universal residue of the photon pole gives the square of the unit
electric charge $\bar{e}^2(0) = 4\pi\alpha$.

A few technical comments are in order.
In eq.(\ref{m_nc}), the matrix elements are linear in the one-loop functions
where the renormalization group improvement is achieved by (\ref{rg}) and
the condition (\ref{rc}).
The use of the running $Z$ width above necessarily modifies the mass
renormalization conditions (\ref{polemass})\cite{mz_pole}, and we adopt
the convention of ref.\cite{mz_lep} for $m_Z$.
The associated small changes in the propagator correction factors
($\Delta_{Z}$ and $\Delta_{\gamma Z}$) are not given explicitly
above for brevity.
The overlines on the vertex functions $\gambar_2^{f_L}(s)$ indicates the
removal of the pinch term\cite{kl,pt2}.
The vertex functions $\Gamma_3^{f_\alpha}(s)$ are proportional to the square
of the fermion mass inside the loop, and are non-vanishing only for
$f_\alpha=b_L$ in the SM.
The functions $\Gamma_4^{f_\alpha}(s)$ are vanishing for all $f_\alpha$
in the SM, though they appear in extended models such as the minimal
SUSY-SM.
The box functions $B_{ij}(s,t)$ are needed only at low energy $NC$ processes
at $s=t=0$.
It is worth noting here that the box contributions to the helicity amplitudes
can be expressed in the above simple current product form only when the
external fermion masses can be neglected.
All the vertex and box functions are known precisely in the SM.
If we assume no new physics contributions to these process specific
($f_\alpha$-dependent) corrections, we can determine the three form factors
$\ebar^2(q^2)$, $\gzbar^2(q^2)$ and $\sbar^2(q^2)$
from the precision experiments independent of further model assumptions.

For the charged current ($CC$) process $ij \to i'j'$, we find similarly
\beq
\label{m_cc}
	M^{CC}_{ij} =\frac{1}{t -m_W^2} \{
	\gwbar^2(t) +\ghat^2[\Gamma_1^{ii'} +\Gamma_1^{jj'}
	+\gambar_2^{ii'} +\gambar_2^{jj'}](t) \}
	+B^{CC}_{ij}(s,t),
\eeq
off the $W$ pole, with an appropriate CKM factor $V_{ii'}V^*_{jj'}$.
Precise values of the $CC$ matrix elements are needed only at low energies,
and we find for the muon decay constant
\beq
\label{gf}
	G_F = \frac{\gwbar^2(0) +\ghat^2\delg}	{4\sqrt{2}m_W^2}.
\eeq
Here the factor $\delg$ denotes the sum of the vertex and the box
contributions, whose value is precisely known ($\delg=0.0055$)
in the SM.
Eq.(\ref{gf}) gives the physical $W$ mass in terms of $G_F$ once
the $\delg$ value is known for a given model.
The overline here again indicates the removal of the pinch terms and that
its numerical value is significantly (about 25\%) smaller than the standard
factor\cite{del_gf}.

In the following, we assume that there are no new physics contributions
to the vertex and box corrections, except that we allow the $Zb_Lb_L$ vertex
to take an arbitrary value, and determine the form factors (\ref{barcharges})
from the three sectors of the electroweak precision experiments.

\subsection*{\normalsize {\bf 3.1 $Z$ boson parameters }}
\pr
The most recent results from experiments at LEP and SLC on the $Z$ boson
parameters have been reported in refs.\cite{zdata93,lep93}.  The $Z$ line-shape
parameters are determined at LEP as\cite{lep93}
\bea
\label{lep93}
   	\left.
      	\begin{array}{lll}
         m_Z({\rm GeV})      & = & 91.187 \pm 0.007  \\
         \Gamma_Z({\rm GeV}) & = &  2.489 \pm 0.007  \\
         \sigma_h^0({\rm nb})& = & 41.56  \pm 0.14   \\
         R_\ell = \sigma_h^0/\sigma_\ell^0
                             & = & 20.763\pm  0.049  \\
         A_{\rm FB}^{0,\ell}     & = & 0.0158 \pm 0.0018
      	\end{array}
   	\right. 
   	\rho_{\rm corr} =
   	\left(
       	\begin{array}{rrrrr}
          1 & -0.157 &  0.017 &  0.012 &  0.075 \\
            &  1     & -0.070 &  0.003 &  0.006 \\
            &        &  1     &  0.137 &  0.003 \\
            &        &        &  1     &  0.008 \\
            &        &        &        &  1
       	\end{array}
   	\right). \hspace{2mm}
\eea
The other electroweak data that we used in our fit are as
follows\cite{zdata93,lep93}:
\bsub
\label{zdata93}
\bea
\label{ptaudata}
  	P_\tau       &=& -0.139  \pm 0.014 ,\\
\label{alrdata}
  	A_{\rm LR}       &=&  0.10   \pm 0.044 \6 \qquad (\mbox{SLD\cite{alr92}}),\\
\label{afbbdata}
  	A_{\rm FB}^{0,b} &=&  0.099  \pm 0.006 ,\\
\label{afbcdata}
  	A_{\rm FB}^{0,c} &=&  0.075  \pm 0.015 ,\\
\label{rbdata}
  	R_b =\sigma_b^0/\sigma_h^0
		     &=&  0.2203 \pm 0.0027 \p3 \quad ({\rm LEP + SLD}).
\eea
\esub
Significant improvements over the last year have been achieved for many
of the above measurements.

In order to determine the universal charge form factors $\sbar^2(m_Z^2)$
and $\gzbar^2(m_Z^2)$ from these data, we should estimate the SM corrections
to the vertex diagrams, QCD higher order effects, and external fermion mass
effects.  We assume that only three neutrinos ($N_\nu = 3$) contribute to
the invisible width of $Z$, and take the standard perturbative QCD
corrections
for the vector\cite{qcd_4l} and axial-vector\cite{qcd_zgg} $Z$ couplings,
the quark mass effects\cite{qcd_mq} and the forward-backward
asymmetries\cite{qcd_afb}.

The fit results are then found to depend on the assumed $\alpha_s$
value and on $m_t$ which affects the $Zb_Lb_L$ vertex function\cite{zbb}.
In the absence of an accurate quantitative measurement
of the QCD coupling constant and for the convenience of the GUT studies,
we choose $\alpha_s \def \alpha_s(m_Z)_{\msbar}$ as a free parameter
of our fit, and present the $\alpha_s$ dependences of the minimal $\chi^2$
values.  One can either add independent data on $\alpha_s$ or study
quantitative consequences of a particular GUT model that predicts $\alpha_s$.

As for the strong $m_t$ dependence of the $Zb_Lb_L$ vertex, we find it
convenient to introduce one extra form factor
\beq
\label{delb}
	\delb(s)
	=\Gamma_1^{b_L}(s) +\chat^2 \gambar_2^{b_L}(s) +\Gamma_3^{b_L}(s)
\eeq
in our fit.  A similar strategy has been proposed in ref.\cite{abc}.
An advantage is that the parameter $\delb$ allows us to determine
the quantitative significance of the $Zb_Lb_L$ vertex correction\cite{zbb_lep},
independent of the specific SM mechanism.
Furthermore, it allows us to separate the data analysis stage from
the evaluation of $\delb$ in a specific model, that includes
$O(\alpha_s m_t^2)$\cite{zbb_qcd} and $O(m_t^4)$\cite{rho_zbb_ew,zbb_ew}
two-loop corrections.

The overall fit to all the $Z$ parameters listed above in terms
of the three parameters $\sbar^2(m_Z^2)$, $\gzbar^2(m_Z^2)$ and
$\delb(m_Z^2)$ for a given value of $\alpha_s$ gives
\bsub
\label{fit_zall}
\bea
   	\left.
      	\begin{array}{lll}
         \gzbar^2(m_Z^2)&\mm =&     0.5546 -0.031(\alpha_s-0.12) \pm 0.0017 \\
         \sbar^2(m_Z^2)	&\mm =&     0.2313 +0.008(\alpha_s-0.12) \pm 0.0007 \\
         \delb(m_Z^2)	&\mm =&\m5 -0.0061 -0.430(\alpha_s-0.12) \pm 0.0035
      	\end{array}
   	\right. 
   	\rho_{\rm corr} =
   	\left(
       	\begin{array}{rrr}
          1 &  0.14 & -0.36 \\
            &  1    &  0.20 \\
	    &       &  1
       	\end{array}
   	\right)
\\
\label{chi_zall}
	\chi^2_{\rm min} =1.60 +((\alpha_s-0.103)/0.0127)^2, \hspace{65mm}
\eea
\esub
where the errors and the correlations are almost independent of $\alpha_s$.

The above results are shown in Fig.~2, along with the SM predictions
with all known corrections to the $\rho$-parameter\cite{rho}
in the $O(m_t^4)$ level\cite{rho_ew1,rho_ew2,rho_zbb_ew,rho_ew3}
and the $O(\alpha_s)$ two-loop corrections\cite{rho_pqcd} in perturbative QCD,
but without non-perturbative $t \bar{t}$ threshold effects\cite{rho_thr}.
The SM prediction to $\sbar^2(m_Z^2)$ is also sensitive to the hadronic
vacuum polarization correction, for which we take\cite{piqq_h}
$(\Delta \frac{1}{\alpha})_{\rm hadrons} =-0.0283/\alpha =-3.88$.
Its error $\delta_\alpha =\pm 0.0007/\alpha =\pm 0.10$ leads
to a shift in the SM predictions for $\sbar^2(m_Z^2)$ by $\pm 0.00026$.

We show in Fig.~2 1-$\sigma$ contours of the fit for three representative
$\alpha_s$ values.  It is clearly seen that the $\gamma Z$-mixing parameter
$\sbar^2(m_Z^2)$ is measured rather insensitively to $\alpha_s$, while
the $Z$ coupling strength $\gzbar^2(m_Z^2)$ is negatively correlated
with the assumed $\alpha_s$ value, reflecting its sensitivity to the
total $Z$ width.  This anti-correlation leads to a preference of larger
$m_t$ in the SM for smaller $\alpha_s$.  The relative insensitivity
of the parameter $\sbar^2(m_Z^2)$ to $\alpha_s$ can easily be understood
since it can be measured from the asymmetry parameters that are either
completely or almost insensitive to the QCD corrections.  We list below
its value determined from each asymmetry measurement:
\bsub
\label{fit_sb2}
\bea
\label{sb2_afbl}
  	\sbar^2(m_Z^2) &=& 0.2309 \pm 0.0010
				\hspace{41mm} ({\rm from}\: A_{\rm FB}^{0,\ell}) ,\\
\label{sb2_ptau}
  	\sbar^2(m_Z^2) &=& 0.2316 \pm 0.0018
				\hspace{41mm} ({\rm from}\: P_{\tau}) ,\\
\label{sb2_alr}
  	\sbar^2(m_Z^2) &=& 0.2365 \pm 0.0055
				\hspace{41mm} ({\rm from}\: A_{\rm LR}) ,\\
\label{sb2_afbb}
  	\sbar^2(m_Z^2) &=& 0.2313 +0.004(\alpha_s-0.12) \pm 0.0011
		        	\quad\:\:  ({\rm from}\: A_{\rm FB}^{0,b}) ,
\eea
\esub
where the first three lepton asymmetries are almost completely insensitive
to $\gzbar^2(m_Z^2)$ or $\alpha_s$, and the $b$-quark forward-backward
asymmetry is also insensitive to $\gzbar^2(m_Z^2)$ or $\delb(m_Z^2)$
while mildly sensitive to $\alpha_s$ due to perturbative QCD
correction\cite{qcd_afb}.  From the above data alone, we find
\bsub
\label{fit2_sb2}
\bea
\label{sb2_lept}
  	\sbar^2(m_Z^2) &=& 0.2312 \pm 0.0009
	\hspace{40mm}	({\rm from}\: A_{\rm FB}^{0,\ell}, P_{\tau}, A_{\rm LR}) ,\\
\label{sb2_asym}
  	\sbar^2(m_Z^2) &=& 0.2312 +0.002(\alpha_s-0.12) \pm 0.0007
	\quad\: ({\rm from}\:
	A_{\rm FB}^{0,\ell}, P_{\tau}, A_{\rm LR}, A_{\rm FB}^{0,b}) .
\eea
\esub
We may expect a significantly improved measurement of $A_{\rm LR}$ from SLD
in the near future\cite{zdata93}.
$\tau$ polarization measurement may still be improved\cite{ptau}.
These asymmetry measurements are particularly important for GUT studies,
since the parameter $\sbar^2(m_Z^2)$ is directly related to the unifying
coupling $\shat^2(\mu)$ via eq.(\ref{rg}).
\begin{center}
\vspace{12cm}
\begin{minipage}{14cm}
{\footnotesize Fig.~2:  Three parameter fits to the $Z$ boson parameters
for three $\alpha_s(m_Z)$ values.  Also shown are and the SM predictions
for $(\Delta\frac{1}{\alpha})_{\rm hadrons} =-3.88$
($\delta_\alpha=0$)\cite{piqq_h}.  }
\end{minipage}
\end{center}

Before leaving the $Z$ parameters, we would like to give two comments
on the measurements of the $Zb_Lb_L$ vertex and $\alpha_s$, which are
strongly correlated.
As is clearly seen from Fig.~2, the fit to the parameter $\delb$ depends
strongly on $\alpha_s$, reflecting its sensitivity to $R_l$ and $\Gamma_Z$,
in addition to $R_b$ that is rather insensitive to $\alpha_s$.  Because of
this sensitivity to $\alpha_s$, it is not meaningful to quote a bound
on $\delb$, or on $m_t$ in the SM, without studying carefully its
$\alpha_s$ dependence.  It is worth emphasizing here that there is
no evidence of the $Zb_Lb_L$ vertex for $\alpha_s \simgt 0.13$, as
the corresponding parameter for $d_L$ or $s_L$ is about $-0.003$.
For $\alpha_s \simgt 0.12$, we can obtain rather stringent upper bound
on $m_t$\cite{zbb_lep,abc} that one can read off from Fig.~2, mainly because
there is no good evidence for the $Zb_Lb_L$ vertex effect.
This point has also been emphasized by the LEP electroweak working
group\cite{lep93}.
Furthermore, this strong correlation makes the fitted $\alpha_s$ value
depend strongly on the assumed $\delb$ value.
If we allow $\delb$ and $\alpha_s$ to be fitted freely by the data, then
the result (\ref{fit_zall}) gives $\delb(m_Z^2) =0.0015\pm0.0071$ and
$\alpha_s(m_Z^2) =0.103\pm 0.013$, with $\rho_{\rm corr}=-0.85$.
It is therefore necessary to assume the SM contributions to $\delb(m_Z^2)$,
and to a lesser extent those to $\gzbar^2(m_Z^2)$,
in order to measure $\alpha_s$ from the electroweak $Z$-parameters.
The result of such an analysis is given in section~4.3 where
we study consequences of the minimal SM.

\subsection*{\normalsize {\bf 3.2 Low energy neutral current experiments }}
\pr
We consider in our analysis four types of low energy neutral current
experiments.  They are the neutrino-nuclei scattering ($\nu_\mu$--$q$),
the neutrino-electron scattering ($\nu_\mu$--$e$), atomic parity violation
(APV), and the polarized electron-deuteron scattering experiments ($e$D).
All of them measure the universal form factors $\sbar^2(0)$ and $\gzbar^2(0)$.
Effects due to small but finite momentum transfer in these processes
are corrected for by assuming that the running of these form factors are
determined by the SM particles only (see Fig.~1), which is an excellent
approximation at low energies.  Vertex and box corrections are performed
by assuming that they are dominated by the SM contributions.
For each sector, we first give a model-independent parametrization of
the data, and then give our fit in the ($\sbar^2(0)$, $\gzbar^2(0)$) plane.

For the $\nu_\mu$--$q$ data, we used the results of the analysis of
ref.\cite{nuq_dat}.  The fitted parameters ($g_L^2$, $g_R^2$, $\delta_L^2$,
$\delta_R^2$) are, however, dependent on the assumed value of the
charmed quark mass ($m_c$) in the slow-rescaling formula for the charged
current cross sections.  By using the constraint on $m_c$ from the charged
current experiments, $m_c =1.54\pm0.33$~GeV\cite{nuq_dat}, we can properly
take into account the $m_c$ dependence of the fit.  We thus find a new
model-independent parametrization of the $\nu_\mu$--$q$ data:
\beq
\label{nuq_dat}
	\left.
	\begin{array}{llr}
	g^2_L 	   &=&  0.2980 \pm 0.0044 \\
	g^2_R 	   &=&  0.0307 \pm 0.0047 \\
	\delta^2_L &=& -0.0589 \pm 0.0237 \\
	\delta^2_R &=&  0.0206 \pm 0.0160
      	\end{array}
   	\right. \quad
   	\rho_{\rm corr} =
   	\left(
       	\begin{array}{rrrr}
          1 &  -0.559 & -0.163 &  0.162 \\
            &   1     &  0.156 & -0.037 \\
	    &         &  1     & -0.447 \\
	    &         &        &  1
       	\end{array}
   	\right).
\eeq
The standard model fit is then performed by expressing the above parameters
in terms of the ratio of the squares of the $NC$ and $CC$ $S$-matrix elements
of eqs.(\ref{m_nc},\ref{m_cc}) evaluated at
$<-t>_{NC} =<-t>_{CC} =20$~GeV$^2$.
We reproduced the well-known results of ref.\cite{nuq_ew}.
The corrections due to the running of $\sbar^2(t)$, the neutrino
'charge radius' factor\cite{nu_chr} $\gambar_2^{\nu_\mu}(t)$ of
eq.(\ref{m_nc}),
and the $WW$ box are found to be significant.
After further correcting for the QED radiation effects in the $CC$
cross section\cite{nuq_cc}, we find
\bsub
\label{fit_nuq}
\bea
   	\left.
      	\begin{array}{lll}
	 \gzbar^2(0) &=& 0.5486 \pm 0.0080 \\
	 \sbar^2(0)  &=& 0.2398 \pm 0.0143
      	\end{array}
   	\right\} \qquad
   	\rho_{\rm corr} =0.92 ,
\\
\label{chi_nuq}
	\chi^2_{\rm min} =0.86 .\hspace{57.5mm}
\eea
\esub
The strong positive correlation is a consequence of the smallness of
the error of $g_L^2 +g_R^2$ in (\ref{nuq_dat}) that measures the total
neutral current cross section off isoscalar targets.
The above fit is given in Fig.~3 as a 1-$\sigma$ counter.

For the $\nu_\mu$--$e$ data, we used the results of CHARM,
BNL~E374 and CHARM-II\cite{nue_dat}, which are summarized by
R.~Beyer\cite{nue_dat} as
\beq
\label{nue_dat}
	\left.
	\begin{array}{llr}
	(\rho)^{\nu_\mu e}_{\rm eff}  &=&  1.007 \pm 0.028 \\
	(\sin^2\theta_W)^{\nu_\mu e}_{\rm eff} &=&  0.233 \pm 0.008
      	\end{array}
   	\right\} \qquad
   	\rho_{\rm corr} = 0.09.
\eeq
These effective parameters are obtained from the data by assuming
the tree-level formula for the $\nu_\mu e$ and $\bar{\nu}_\mu e$
scattering cross sections.
We can hence obtain the electroweak parameters by evaluating
the full matrix elements at an average momentum transfer of these experiments,
$<-t> \sim m_\mu^2$.
We reproduce the known results of ref.\cite{nue_ew}, and find
that the only significant correction comes from the neutrino
'charge radius' factor and the $WW$ box contributions.  We find
\beq
\label{fit_nue}
   	\left.
      	\begin{array}{lll}
         \gzbar^2(0)	&=&     0.5459 \pm 0.0153 \\
         \sbar^2(0)	&=&     0.2416 \pm 0.0080
      	\end{array}
   	\right\} \qquad
   	\rho_{\rm corr} =0.09
\eeq
with $\chi^2_{\rm min} =0$, since we take the fit (\ref{nue_dat})
as the model independent parametrization of the $\nu_\mu -e$
data\cite{nue_dat}.
The result is also shown in Fig.~3.

As for the APV experiments, we used the result of the analysis\cite{apv_dat}
on the parity violating transitions in the cesium atom (A,Z)=(135,55);
\beq
\label{apv_dat}
	Q_W(135,55) = -71.04 \pm 1.81
\eeq
where we sum the experimental and theoretical errors by quadrature.
Our simple formula (\ref{m_nc}) reproduces the $u$- and $d$-quark contributions
of ref.\cite{apv_ew}, but not the photonic correction to the axial vector
$Zee$ vertex nor the $Z\gamma$ box corrections that are sensitive to the
nucleon structure.  We adopt the results of ref.\cite{apv_ew} for these
corrections, and find
\beq
\label{fit_apv}
	\sbar^2(0) = -0.6130 \cdot\gzbar^2(0) +0.5661 \pm 0.0083
\eeq
The result is shown in Fig.~3.

Finally, for the SLAC $e$D polarization asymmetry experiment\cite{ed_dat},
we make a model-independent fit to the original data by using the two
parameters, $2C_{1u}-C_{1d}$ and $2C_{2u}-C_{2d}$ of ref.\cite{old_ncfit},
by taking into account uncertainties due to the sea-quark contributions
and finite $R=\sigma_L/\sigma_T$\cite{ed_qcd}, and those due to
higher twist contributions\cite{ed_ht1,ed_ht2}.
The former uncertainties are found to be very small, confirming the
results of ref.\cite{ed_qcd}, while the latter are found to be model
dependent\cite{nuq_ht2}.
We adopt the estimates\cite{ed_ht2} based on the MIT-Bag model,
which find rather small corrections, as in the neutrino scattering
off isoscalar targets\cite{nuq_ht1}.
Further study on the higher twist effects may be needed to achieve
precision measurements of the electroweak parameters in these reactions.
After allowing for uncertainties in the Bag model parameters of
ref.\cite{ed_ht2}, we find
\beq
\label{ed_dat}
   	\left.
      	\begin{array}{llr}
         2C_{1u}-C_{1d} &=& -0.94 \pm 0.26 \\
         2C_{2u}-C_{2d} &=&  0.66 \pm 1.23
      	\end{array}
   	\right\} \qquad
   	\rho_{\rm corr} =-0.975
\eeq
with $\chi^2_{\rm min}=9.95$ for 11 data points.  Because of the strong
correlation, only a linear combination of the two coupling factors is measured
well.  The electroweak corrections in the SM are found in ref.\cite{ed_ew}.
Our formula (\ref{m_nc}) leads to all relevant correction factors
except for the external photonic corrections.  We use the explicit
form of ref.\cite{apv_ew} for these correction factors, and checked
the insensitivity of our fit to the uncertainty in the $Z\gamma$ box
corrections.
The QED coupling $\ebar^2(t)$ and the vertex functions $\Gamma_1(t)$ and
$\gambar_2(t)$ in our amplitudes (\ref{m_nc}) are evaluated at
$<-t>=1.5$GeV$^2$.  We find
\bsub
\label{fit_ed}
\bea
	\sbar^2(0) &=& 0.3264\cdot\gzbar^2(0) +0.0471 \pm 0.0094
\\
\label{chi_ed}
	\chi^2_{\rm min} &=& -1.77 \cdot\gzbar^2(0) +1.43,
\eea
\esub
where $\chi^2_{\rm min}$ is obtained by taking our model-independent fit
(\ref{ed_dat}) as our input 'data'.

The results of our two parameter fit to all the neutral current data are
summarized in Fig.~3 by 1-$\sigma$ allowed regions in the
($\sbar^2(0),\;\gzbar^2(0)$) plane.
They are consistent with each other and, after combining the above four
sectors, we find
\bsub
\label{fit_lenc}
\bea
   	\left.
      	\begin{array}{lll}
         \gzbar^2(0)	&=&     0.5459 \pm 0.0035 \\
         \sbar^2(0)	&=&     0.2350 \pm 0.0045
      	\end{array}
   	\right\} \qquad
   	\rho_{\rm corr} =0.53,
\\
\label{chi_lenc}
	\chi^2_{\rm min} = 2.82. \hspace{58mm}
\eea
\esub
The fit is excellent as the effective degrees of freedom of the fit
is $8-2=6$.  The combined fit above is shown by the thick 1-$\sigma$ contour
in Fig.~3.
\begin{center}
\vspace{6cm}
{\footnotesize Fig.~3:  Two parameter fit to L.E.N.C. data.
\hspace{10mm} Fig.~4:  $Z$ parameter fit plus L.E.N.C. fit. \hspace{10mm} }
\end{center}

\subsection*{\normalsize {\bf 3.3 Charged current experiments }}
\pr
The $W$ mass data have been updated this summer by the CDF and D0
collaborations\cite{mw93}.  We obtain
\beq
\label{mw_dat}
	m_W = 80.25 \pm 0.24 {\rm GeV}
\eeq
by combining the two most recent measurements\cite{mw93} after
adding all the quoted errors by quadrature.

The electroweak parameter $\gwbar^2(0)$ is then obtained from the
$\mu$ life-time via the identity (\ref{gf}).
By using the SM estimate $\delg =0.0055$ and the perturbative approximation
$\ghat^2 =\gwbar^2(0)$, we find
\beq
\label{fit_mw}
	\gwbar^2(0) = 0.4226 \pm 0.0025.
\eeq
No other experiment in the charged current sector is accurate enough
to add useful information in our electroweak analysis.
Precision measurements of the $W$ width\cite{wwidth_ew} and its
leptonic branching fraction may determine $\gwbar^2(m_W^2)$
in the future.

\section*{\normalsize {\bf 4.
SYSTEMATIC ANALYSIS }}
\pr
All the electroweak precision data have now been represented by the
charge form factor values of eqs.(\ref{fit_zall},\ref{fit_lenc},\ref{fit_mw}).
We find that all results are consistent with the assumptions
of the ${\rm SU(2)_L \times U(1)_Y}$ universality and the SM dominance of the
vertex and box corrections.
In the following, we perform the fit to the data in three steps by
systematically strengthening the model assumptions.

\subsection*{\normalsize {\bf 4.1
Testing the running of the charge form factors }}
\pr
Only two of the four form factors, $\sbar^2(q^2)$ and $\gzbar^2(q^2)$,
have been measured sufficiently accurately at two energy scales,
$q^2 = 0$ and $m_Z^2$.  From eqs.(\ref{fit_zall},\ref{fit_lenc}),
we find
\beq
\label{run_fit}
   	\left.
      	\begin{array}{lll}
	\gzbar^2(m_Z^2) -\gzbar^2(0) &=& \3h
	 0.0087 -0.031(\alpha_s-0.12) \pm 0.0039 \\
	\sbar^2(m_Z^2) -\sbar^2(0) &=&
	-0.0037 +0.008(\alpha_s-0.12) \pm 0.0046
      	\end{array}
   	\right\} \qquad
   	\rho_{\rm corr} =0.48.
\eeq
The SM predictions for these quantities are, respectively,
\bsub
\label{run_sm}
\bea
	{[\gzbar^2(m_Z^2) -\gzbar^2(0)]}_{\rm SM} &=& \3h 0.00723
	+10^{-4}[2.5(\frac{m_t}{150}-1) -0.15\ln\frac{m_H}{100}
	-0.56(\frac{100}{m_H})^2], \hspace{6mm}
\\
	{[\sbar^2(m_Z^2) -\sbar^2(0)]}_{\rm SM}	&=& -0.00838
	-10^{-4}[3.1(\frac{m_t}{150}-1) -0.33\ln\frac{m_H}{100}],
\eea
\esub
where $m_t$ and $m_H$ are measured in GeV units.
Both results are consistent at the 1-$\sigma$ level with the assumption
that the running of these form factors is governed by the SM particles
only.  Since the running of the form factors is affected only by
particles of mass in the vicinity of $m_Z$, we conclude that there is
no indication of new particles of mass $\simlt m_Z$.

The errors in (\ref{run_fit}) are determined by those of the low energy
experiments.
Further improvements in the low energy precision experiments are needed
to detect a signal of relatively light new particles.

\subsection*{\normalsize {\bf 4.2
Testing the three parameter universality }}
\pr
Once we assume further that the running of the charge form factors is
governed by the SM particles only, then we can parametrize all the predictions
of the general ${\rm SU(2)_L \times U(1)_Y}$ model in terms of just three free
parameters; see Table~1.
This can easily be understood by noting that there are
6 parameters in the gauge boson sector of the model;
the four charge form factors associated with the four gauge boson propagators
and the two gauge boson masses $m_W$ and $m_Z$.
{}From these 6 parameters, 3 should be traded for the three fundamental
parameters of the theory, the two gauge couplings and one vacuum expectation
value.  We choose for convenience the three most accurately measured
quantities, $\alpha$, $G_F$ and $m_Z$, as the parameters which
renormalize the sensitivity to physics at very high energies.
The remaining 3 parameters can hence reveal one-loop physics at the
weak scale.  We first choose $\sbar^2(m_Z^2)$, $\gzbar^2(m_Z^2)$
and $\gwbar^2(0)$ as the three parameters, and present the result
of our global analysis.  The result is then re-expressed in terms of
another set of the parameters, $S$, $T$ and $U$\cite{stu}.

By using the SM running of the form factors (\ref{run_sm}), we can combine
the $Z$ parameter fit (\ref{fit_zall}) and the low energy $NC$ fit
(\ref{fit_lenc}).  This is schematically shown in Fig.~4, where the
combined low energy $NC$ fit of Fig.~3 is reproduced in the
($\sbar^2(m_Z^2)$, $\gzbar^2(m_Z^2)$) plane.  The uncertainty
in the running of the parameters within the SM is visualized by
the thickness of the contour which spans the range $m_t = 100 - 200$~GeV,
$m_H = 100 - 1000$~GeV in eq.(\ref{run_sm}).  The low energy parameters
are consistent with the $Z$ parameters, which are also shown as
the 'LEP$+$SLC' contour.  All the neutral current data are now combined
to give
\bsub
\label{fit_allnc}
\bea
   	\left.
      	\begin{array}{llr}
	\gzbar^2(m_Z^2)\mm &=& \mm 0.5547 -0.023(\alpha_s-0.12) \pm 0.0015 \\
	\sbar^2(m_Z^2) \mm &=& \mm 0.2312 +0.008(\alpha_s-0.12) \pm 0.0007 \\
	\delb(m_Z^2)   \mm &=& \mm -0.0063 -0.437(\alpha_s-0.12) \pm 0.0034
      	\end{array}
   	\right. \:\:
   	\rho_{\rm corr} =
   	\left(
       	\begin{array}{rrr}
          1 &   0.16 & -0.31 \\
            &   1    &  0.20 \\
	    &        &  1
       	\end{array}
   	\right), \hspace{4mm} \\
\label{chi_allnc}
	\chi^2_{\rm min} =5.40 +((\alpha_s-0.103)/0.0123)^2. \hspace{75mm}
\eea
\esub
The above fit is almost independent of ($m_t$, $m_H$) values assumed in
the running of the charge form factors.
The $\chi^2_{\rm min}$ value of 7.3 for $\alpha_s=0.12$ is excellent
for the effective degrees of freedom of the fit, $18-3=15$.

There is one notable point at this stage which becomes apparent by comparing
the global fit of Fig.~4 with the individual fit to low energy $NC$ data
in Fig.~3.  Both the data on $\nu_\mu$--$q$ and $\nu_\mu$--$e$ experiments
are perfectly consistent with the global fit, whereas the APV result and
the $e$D asymmetry fit are just 1-$\sigma$ away.
Further studies of polarization asymmetries in the $e-q$ sector,
as well as quantitative studies of the neutral current processes
at TRISTAN energies might be potentially rewarding.

The global fit (\ref{fit_allnc}) for ($\sbar^2(m_Z^2)$, $\gzbar^2(m_Z^2)$)
from all the neutral current data and the fit (\ref{fit_mw}) for $\gwbar^2(0)$
from the charged current data summarize our knowledge on the electroweak
parameters in our framework; see Table~1.

When the basic three parameters of the models with the
${\rm SU(2)_L \times U(1)_Y}$ symmetry broken by just one vacuum expectation
value are renormalized by the three well-known quantities $\alpha$, $G_F$ and
$m_Z$, all the predictions of the theory are determined at the tree level.
It is therefore convenient to introduce three parameters which are
proportional to the finite quantum correction effects only.
Among the various proposals in the literature\cite{stu,stu_mod,ab}, we find
that the $S$, $T$, $U$ parameters of Peskin and Takeuchi\cite{stu} is most
convenient if they are extended to include the SM contributions as well.
We {\it define} these parameters in terms of our two-point functions with
the pinch terms\cite{pt2}, which are related to our charge form factors
as follows:
\bsub
\label{stu}
\bea
\label{s}
	S &\equiv& \frac{1}{\pi}
	Re[\pibar^{3Q}_{T,\gamma}(m_Z^2) -\pibar^{33}_{T,Z}(0)]
	= \frac{4\sbar^2(m_Z^2)\cbar^2(m_Z^2)}
	{\bar{\alpha}(m_Z^2)} -\frac{16\pi}{\gzbar^2(0)} ,\\
\label{t}
	\alpha T &\equiv& \frac{G_F}{2\sqrt{2}\pi^2} \:
	[\pibar^{33}_T(0) -\pibar^{11}_T(0)] \:\:\:
	= 1 +\delg -\frac{4\sqrt{2}G_F m_Z^2}{\gzbar^2(0)} ,\\
\label{u}
	S+U &\equiv& \frac{1}{\pi}
	Re[\pibar^{3Q}_{T,\gamma}(m_Z^2) -\pibar^{11}_{T,Z}(0)]
	= \frac{4\sbar^2(m_Z^2)}{\bar{\alpha}(m_Z^2)}
	-\frac{16\pi}{\gwbar^2(0)}.
\eea
\esub
These definitions allow us to express all the charge form factors and
hence all experimental observables in terms of the three parameters
$S$, $T$ and $U$ without separating the SM contributions to the
gauge boson propagators.
First, the form factor $\gzbar^2(0)$ is determined from $T$ via eq.(\ref{t}).
Second, the form factor $\sbar^2(m_Z^2)$ is determined from $S$ via
eq.(\ref{s}).  And finally the form factor $\gwbar^2(0)$ is determined
from $U$ via eq.(\ref{u}).
The running of these form factors is determined by their defining equations
(\ref{barcharges}) by properly performing the renormalization group
improvement via eq.(\ref{rg}).
All the form factors are thus easily calculable for arbitrary models
for fixed ($\alpha$, $G_F$ and $m_Z$).
In fact, the SM curves in Figs.~1 and 2 are obtained this way.

By assuming only the SM contribution to the muon decay vertex and box
corrections in $\delg$ and by assuming the SM running of the form factors,
especially for $\bar{\alpha}(q^2)$,
we can express all the charge form factors as a power series in the above
three parameters.  To first order, we find
\begin{subequations}
\bea
	\gzbar^2(0)    &=& 0.5456 \hphantom{ +0.0036S }\;\, +0.0040T  ,\\
	\sbar^2(m_Z^2) &=& 0.2334            +0.0036S  -0.0024T
	   \hphantom{ +0.0035U }\;\, -0.0026\delta_\alpha	 ,\\
	\gwbar^2(0)    &=& 0.4183            -0.0031S  +0.0044T +0.0035U
                                      -0.0015\delta_\alpha	,
\eea
\esub
where we added the shifts due to the uncertainty in the estimated
$1/\bar{\alpha}(m_Z^2)$ value, $\delta_\alpha$, which is as large as
$\pm 0.10$\cite{piqq_h}.
It is clearly seen that $\gzbar^2(0)$ measures $T$, $\sbar^2(m_Z^2)$
measures a combination of $S$ and $T$, whereas $\gwbar^2(0)$ measures
a combination of all three parameters.

We can now express the result of our global fit (\ref{fit_allnc}) and
(\ref{fit_mw}), in terms of the $S$, $T$, $U$ parameters and $\delb(m_Z^2)$.
We find
$$
	\left.
	\begin{array}{lll}
	S  &=& \mm\mm -0.29 -1.6(\alpha_s-0.12)
			+0.73\delta_\alpha \pm 0.33 \\
	T  &=&         0.46 -5.8(\alpha_s-0.12)
			-0.04\delta_\alpha \pm 0.37 \\
	U  &=&         0.39 +5.8(\alpha_s-0.12)
			+0.25\delta_\alpha \pm 0.76 \\
	\delb \mm &=& \mm\mm -0.0063 -0.44(\alpha_s-0.12)  \pm 0.0034
      	\end{array}
   	\right.
   	\rho_{\rm corr} =
   	\left(
       	\begin{array}{rrrr}
          1 &  -0.83 & -0.13 & -0.12 \\
            &   1    & -0.29 & -0.31 \\
	    &        &  1    &  0.14 \\
	    &        &       &  1
       	\end{array}
   	\right), \hspace{27mm}
$$
\vglue-11mm
\bsub
\bea
\label{fit_stu}
\\
\label{chi_stu}
	\chi^2_{\rm min} =5.40 +((\alpha_s-0.103)/0.0124)^2
				+(\delta_\alpha/0.1)^2. \hspace{69mm}
\eea
\esub
Only the correlation between the errors in $S$ and $T$ is significant.
We show in Fig.~5 the above results on the ($S$, $T$) and ($U$, $T$) planes.
The only radiative effect which is significantly non-vanishing is
in the $T$ parameter.  Both the $S$ and $U$ parameters are
consistent with zero at the 1-$\sigma$ level.
Note also that the $S$ parameter is particularly sensitive to the hadronic
uncertainty $\delta_\alpha$ of $1/\bar{\alpha}(m_Z^2)$, whose magnitude
can change by a quarter for $\delta_\alpha =\pm0.10$\cite{piqq_h}.
\begin{center}
\vspace{6cm}
\baselineskip4pt
\begin{minipage}{14cm}
{\footnotesize
Fig.~5:  Global fit to the ($S$, $T$, $U$) parameters as
defined in eq.(\ref{stu}) for three $\alpha_s$ values and for ($m_t$, $m_H$)
values in the range $m_t =100 -200$~GeV and $m_H =100-1000$~GeV,
for $(\Delta\frac{1}{\alpha})_{\rm hadron}=-0.0283/\alpha =-3.88$
($\delta_\alpha=0$)\cite{piqq_h}.
The SM predictions are also given.  }
\end{minipage}
\end{center}
\baselineskip14pt

\subsection*{\normalsize {\bf 4.3
Testing the minimal standard model }}
\pr
In the minimal SM, the three parameters $S$, $T$, $U$ as defined above
and the $Zb_Lb_L$ vertex form factor $\delb(m_Z^2)$ are uniquely determined
in the one-loop level by the two mass parameters $m_t$ and $m_H$.
We show in Fig.~6 the SM predictions for these parameters as functions
of $m_t$ for selected values of $m_H$, by including all the known two-loop
corrections of $O(m_t^4)$\cite{rho_zbb_ew,rho_ew3,zbb_ew}
and of $O(\alpha_s)$\cite{rho_pqcd,zbb_qcd} at $\alpha_s(m_Z) =0.12$.
{}From Figs.~5 and 6, one can see that the parameters $S$ and $T$ show mild
sensitivity to $m_H$, but the parameters $U$ and $\delb(m_Z^2)$ are
almost independent of $m_H$.
\begin{center}
\vspace{12cm}
\begin{minipage}{14cm}
{\footnotesize Fig.~6:  The SM predictions for the ($S$, $T$, $U$, $\delb$)
parameters as defined in eqs.(\ref{stu},\ref{delb}) as functions of $m_t$
for selected $m_H$ values.  We set $\alpha_s=0.12$ in the two-loop
$O(\alpha_s)$ corrections for $S$, $T$, $U$\cite{rho_pqcd} and
$\delb(m_Z^2)$\cite{zbb_qcd}.	}
\end{minipage}
\end{center}

By inserting these SM ($m_t$, $m_H$) dependences into our global fits
(\ref{fit_zall},\ref{fit_lenc},\ref{fit_mw}), we find
an excellent agreement of the data with the SM.  In other words, we
find no signal of new physics beyond the SM in the present precision
experiments.

In Fig.~7, we show the result of our global SM fit to all the electroweak
data in the ($m_t$, $m_H$) plane for three representative $\alpha_s$
values.  One can clearly see the positive correlation between the
preferred values of $m_t$ and $m_H$, which is found independently of
the assumed $\alpha_s$ value.  On the other hand, the preferred range of
$m_H$ depends rather sensitively on $\alpha_s$.
For $\alpha_s(m_Z) \simlt 0.125$, smaller $m_H$ is preferred, whereas
for $\alpha_s(m_Z) \simgt 0.130$, larger $m_H$ is slightly favored.
The $m_H$ dependence of the fit is very mild and no strict bound on
$m_H$ can be given without imposing a constraint on $\alpha_s(m_Z)$.
We find e.g. for a relatively small $\alpha_s$ estimate of PDG\cite{pdg92};
\beq
\label{fit_mh}
	\chi^2_{\rm min} = 7.3 +0.25[\ln(m_H/24.6)]^2
			\4\4 \mbox{for}\:\: \alpha_s(m_Z)=0.1134 \pm0.0035.
\eeq
If we blindly take the above $m_H$ dependence of the fit in the region
60~GeV$<m_H<\infty$, we find $m_H<1.0$~TeV (90\% C.L.), confirming the trend
as observed in refs.\cite{ellisfogli,mh_bound}.
\begin{center}
\vspace{7cm}
\begin{minipage}{14cm}
{\footnotesize Fig.~7:  Electroweak constraints on ($m_t$, $m_H$) in
the minimal SM, for three selected $\alpha_s$ and at $\delta_\alpha=0$. }
\end{minipage}
\end{center}

Instead, we may allow the electroweak data alone to constrain $\alpha_s(m_Z)$
as well, extending the analysis of the LEP electroweak working
group\cite{lep93}.
We find the following parametrization of our global fit to all the
electroweak data (\ref{fit_zall},\ref{fit_lenc},\ref{fit_mw}) in terms of
the three parameters $(m_t, m_H, \alpha_s(m_Z))$ in the minimal SM:
\bsub
\label{fit_all_sm}
\bea
	m_t &=& 147 -3(\frac{\alpha_s-0.12}{0.01})
		    -5(\frac{\delta_\alpha}{0.10})
		    +12.8(\ln\frac{m_H}{100})
		    +0.9(\ln\frac{m_H}{100})^2
\nonumber\\
	&& \mbox{\hspace{82mm}}	\left\{
	\begin{array}{l}
	+[15 -0.7\ln(m_H/100)] \\
	-[17 -0.9\ln(m_H/100)]
      	\end{array}
   	\right.
\\
\label{chi_all_sm}
	\chi^2_{\rm min} &=& 7.2 +(\frac{\alpha_s-0.117}{0.0067})^2
	+(1.77-13.0\alpha_s)[\ln\frac{m_H}{23.1}
			-(\frac{\alpha_s-0.101}{0.021})^4]^2
	+(\frac{\delta_\alpha}{0.10})^2. \hspace{10mm}
\eea
\esub
Here $m_t$ and $m_H$ are measured in GeV units.  The parametrization
reproduces the correct $\chi^2_{\rm min}$ within a few \% accuracy in the
range $0.11<\alpha_s(m_Z)<0.13$ and $60<m_H({\rm GeV})<1000$.
A rather complicated functional form of $\chi^2_{\rm min}$ above gives
the aforementioned $\alpha_s$ dependence of the preferred $m_H$ range.
If we allow $m_t$ to take arbitrary values in the range,
$100<m_t(\rm GeV)<200$, then the above fit gives
\beq
\label{fit_alps}
	\alpha_s(m_Z)_{\msbar} = 0.118 +0.0018\ln(m_H/100) \pm 0.006.
\eeq
The mean value above is, however, also sensitive to $m_t$.
We may further allow $m_t$ and $\alpha_s$ to be freely fitted by the
above electroweak data, and find $\chi^2_{\rm min}$ for a given $m_H$:
\beq
\label{chi_all_mh}
	\chi^2_{\rm min} = 6.9 +0.114[\ln(m_H/13.5)]^2
			\4\4 \mbox{for free $\alpha_s(m_Z)$}.
\eeq
Again in the region 60~GeV$<m_H<\infty$, this leads to a formal constraint
on $m_H$: $m_H < 3.4~{\rm TeV} \:\:\: (90\% C.L.)$.
The upper bound is, however, clearly outside the region of validity of
our perturbative framework.

Finally, we show in Table~2 the complete list of all the input
data (except for $\alpha$, $G_F$ and $m_Z$) and the corresponding minimal
SM predictions for several sets of $(m_t, m_H, \alpha_s)$ values.
The total $\chi^2$ of each sector is also shown in the table,
which is obtained by properly taking account of the correlations among
the errors which are all given in the text.
We see clearly from the table that the present electroweak experiments are
consistent with the SM, perhaps except for a combination of a heavy top and
a light Higgs; see the $(m_t,m_H)=(200,100)$GeV column in the table.
Even there, the total $\chi^2$ over the effective number of degrees of
freedom, 18, is only 1.2.
We find the table very useful, nevertheless, because
we can read off from it the significance of the future improvements
in the precision experiments.
\begin{center}
{\footnotesize Table~2}\\
\end{center}
\begin{center}
 \def\equnit{$\times 10^{-5}$}
 \def\afb{A_{\rm FB}}
 \def\non{$\;\;$------}
 \scriptsize

 \begin{tabular}{|l|r|l|l|llllll|}
 \hline
  & data \hspace{7mm} & no-EW &
  \hspace{2mm}IBA  & & &
  \hspace{4mm}  SM & & & \\
 \hline
         $m_t$ (GeV)   &                &
 \non   &    150 &    150 &    150 &    200 &    200 &    120 &    200\\
         $m_H$ (GeV)   &                &
 \non   &    100 &    100 &   1000 &    100 &   1000 &     60 &   1000\\
        $\alpha_s(m_Z)$&                &
  0.120 &  0.120 &  0.120 &  0.120 &  0.120 &  0.120 &  0.110 &  0.130\\
 \hline
$S$                    &                &
 \non   &-0.2115 &-0.2115 &-0.0540 &-0.2478 &-0.0903 &-0.2332 &-0.0892\\
$T$                    &                &
 \non   & 0.5852 & 0.5852 & 0.3002 & 1.2322 & 0.9110 & 0.3144 & 0.8960\\
$U$                    &                &
 \non   & 0.3032 & 0.3032 & 0.2984 & 0.4103 & 0.4055 & 0.2136 & 0.4045\\
$\bar{\delta}_G$       &                &
 \non   & 0.0055 & 0.0055 & 0.0055 & 0.0055 & 0.0055 & 0.0055 & 0.0055\\
$1/\bar{\alpha}(m_Z^2)$&                &
 128.85 & 128.72 & 128.72 & 128.72 & 128.71 & 128.71 & 128.73 & 128.71\\
$\bar{s}^2(m_Z^2)$     &                &
 0.2312 & 0.2312 & 0.2312 & 0.2325 & 0.2295 & 0.2309 & 0.2317 & 0.2309\\
$\bar{g}_Z^2(m_Z^2)$   &                &
 0.5486 & 0.5552 & 0.5552 & 0.5540 & 0.5578 & 0.5566 & 0.5539 & 0.5565\\
$\bar{\delta}_b(m_Z^2)$&                &
 \non   & \non   &-0.0079 &-0.0079 &-0.0123 &-0.0123 &-0.0058 &-0.0122\\
$\bar{s}^2(0)$         &                &
 0.2388 & 0.2396 & 0.2396 & 0.2408 & 0.2380 & 0.2393 & 0.2401 & 0.2393\\
$\bar{g}_Z^2(0)$       &                &
 0.5486 & 0.5480 & 0.5480 & 0.5468 & 0.5506 & 0.5493 & 0.5469 & 0.5492\\
$\bar{g}_W^2(0)$       &                &
 0.4218 & 0.4226 & 0.4226 & 0.4208 & 0.4259 & 0.4240 & 0.4212 & 0.4239\\
 \hline
$\Gamma_Z$(GeV) &   2.489 $\pm$  0.007  &
  2.485 &  2.514 &  2.490 &  2.482 &  2.503 &  2.494 &  2.480 &  2.499\\
$\sigma_h^0$(nb)&   41.56 $\pm$   0.14  &
  41.47 &  41.47 &  41.45 &  41.46 &  41.48 &  41.49 &  41.49 &  41.43\\
$R_\ell$        &  20.763 $\pm$  0.049  &
 20.807 & 20.807 & 20.763 & 20.740 & 20.745 & 20.722 & 20.708 & 20.789\\
$\afb^{0,\ell}$ &  0.0158 $\pm$ 0.0018  &
 0.0167 & 0.0168 & 0.0153 & 0.0132 & 0.0182 & 0.0159 & 0.0144 & 0.0158\\
$P_\tau$        &  -0.139 $\pm$  0.014  &
 -0.149 & -0.150 & -0.142 & -0.132 & -0.155 & -0.144 & -0.137 & -0.144\\
$A_{LR}$        &   0.100 $\pm$  0.044  &
  0.149 &  0.150 &  0.142 &  0.132 &  0.155 &  0.144 &  0.137 &  0.144\\
$R_b$           &  0.2203 $\pm$ 0.0027  &
 0.2185 & 0.2185 & 0.2165 & 0.2166 & 0.2147 & 0.2147 & 0.2174 & 0.2147\\
$\afb^{0,b}$    &   0.099 $\pm$  0.006  &
  0.105 &  0.105 &  0.099 &  0.092 &  0.109 &  0.101 &  0.096 &  0.101\\
$\afb^{0,c}$    &   0.075 $\pm$  0.015  &
  0.075 &  0.075 &  0.071 &  0.065 &  0.078 &  0.072 &  0.068 &  0.072\\
$\chi^2$        &                       &
   5.35 &  18.01 &   3.71 &   8.28 &  15.89 &   6.91 &   6.24 &   8.73\\
 \hline
$g_L^2$         &  0.2980 $\pm$ 0.0044  &
 0.2954 & 0.2941 & 0.2979 & 0.2958 & 0.3018 & 0.2995 & 0.2963 & 0.2994\\
$g_R^2$         &  0.0307 $\pm$ 0.0047  &
 0.0308 & 0.0309 & 0.0296 & 0.0298 & 0.0295 & 0.0297 & 0.0296 & 0.0297\\
$\delta_L^2$    & -0.0589 $\pm$ 0.0237  &
-0.0600 &-0.0600 &-0.0761 &-0.0760 &-0.0765 &-0.0764 &-0.0759 &-0.0764\\
$\delta_R^2$    &  0.0206 $\pm$ 0.0160  &
 0.0185 & 0.0185 & 0.0177 & 0.0178 & 0.0177 & 0.0178 & 0.0177 & 0.0178\\
$\chi^2 $       &                       &
   0.51 &   1.09 &   0.89 &   1.37 &   1.72 &   0.95 &   1.20 &   0.94\\
 \hline
$s^2_{eff}$     &   0.233 $\pm$  0.008  &
  0.239 &  0.239 &  0.231 &  0.233 &  0.231 &  0.232 &  0.233 &  0.232\\
$\rho_{eff}$    &   1.007 $\pm$  0.028  &
  1.000 &  0.999 &  1.011 &  1.009 &  1.016 &  1.013 &  1.009 &  1.013\\
$\chi^2$        &                       &
   0.59 &   0.77 &   0.09 &   0.01 &   0.21 &   0.07 &   0.01 &   0.07\\
 \hline
$Q_W$           &  -71.04 $\pm$   1.81  &
 -75.49 & -75.57 & -73.12 & -73.22 & -73.13 & -73.23 & -73.09 & -73.23\\
$\chi^2$        &                       &
   6.04 &   6.27 &   1.32 &   1.46 &   1.34 &   1.47 &   1.28 &   1.46\\
 \hline
$2C_{1u}-C_{1d}$&   -0.94 $\pm$   0.26  &
  -0.71 &  -0.71 &  -0.72 &  -0.71 &  -0.72 &  -0.72 &  -0.71 &  -0.72\\
$2C_{2u}-C_{2d}$&    0.66 $\pm$   1.23  &
  -0.08 &  -0.07 &  -0.10 &  -0.09 &  -0.11 &  -0.10 &  -0.09 &  -0.10\\
$\chi^2$        &                       &
   1.95 &   2.15 &   1.56 &   1.83 &   1.21 &   1.45 &   1.71 &   1.46\\
 \hline
$m_W$           &   80.25 $\pm$   0.24  &
  79.95 &  80.25 &  80.25 &  80.08 &  80.57 &  80.38 &  80.11 &  80.38\\
$\chi^2$        &                       &
   1.53 &   0.00 &   0.00 &   0.49 &   1.74 &   0.31 &   0.33 &   0.28\\
 \hline
  $\chi^2_{\rm tot}$&                   &
  15.97 &  28.29 &   7.58 &  13.43 &  22.10 &  11.17 &  10.76 &  12.94\\
 \hline
 \end{tabular}
\end{center}

In Table~2, we also show the results of two approximations:
the 'no-EW' column is obtained by dropping all electroweak corrections
to the two-point functions ($S=T=U=0$) and vertex/box corrections
($\delg=\delb=\Gamma_i=B_{ij}=0$), while retaining the QED running of
the charge form factors $\ebar^2(q^2)$ and $\sbar^2(q^2)$ due to
light particles (excluding $W$ and $t$ contributions).
The 'IBA' column shows the result of the improved Born approximation,
where we retain all the gauge boson propagator corrections and hence
keep all the four charge form factors exact but drop all vertex/box
corrections ($\delb=\Gamma_i=B_{ij}=0$), except for $\delg$ in the
$\mu$ decay.
It is quite surprising to note that the 'no-EW' fit to all the data is
almost as good as the full SM fit for a preferred ($m_t$, $m_H$) range,
and that it is significantly {\it better} than the 'IBA' fit in which all
the most important electroweak corrections are supposed to be contained,
including the dominant $m_t^2$ corrections in the $T$ parameter.
Even more strikingly, if we further set $\delg=0$ in IBA, which
may be called a genuine IBA, we obtain $\sbar^2(m_Z^2)=0.2293$ and
the total $\chi^2$ jumps to 71.
This confirms the observation of ref.\cite{okun} that there is no
evidence of the genuine electroweak correction in the present electroweak
precision experiments, because of the accidental cancellation between
the propagator corrections and the remaining vertex/box corrections.
Strictly speaking, we do not yet have a 'clean evidence'\cite{hioki}
of the genuine electroweak radiative effect.

\section*{\normalsize {\bf 5. CONCLUSIONS }}
\pr
We reported the result of our new global study~\cite{hhkm}
of the electroweak precision measurements.
It introduces four charge form factors
$\ebar^2(q^2)$, $\sbar^2(q^2)$, $\gzbar^2(q^2)$ and $\gwbar^2(q^2)$
associated with the four gauge boson propagators ($\gamma\gamma$,
$\gamma Z$, $ZZ$ and $WW$) of the ${\rm SU(2)_L \times U(1)_Y}$ models.
By assuming negligible new physics contributions to vertex and box
corrections, we can determine these charge form factors accurately
from precision experiments at the one-loop level.
Our approach allows us to test the electroweak theory at several
qualitatively different levels:  first, the ${\rm SU(2)_L \times U(1)_Y}$
universality can be tested by taking all the four charge form factors
to be free parameters; second, the running of the form factors can be
tested against the expectations of the SM; and, third, the normalization
of the three form factors $\sbar^2(m_Z^2)$, $\gzbar^2(m_Z^2)$,
$\gwbar^2(0)$ can be tested against the prediction of the minimal SM.
The data show excellent agreement with the SM at all stages of the
above tests.

We clearly need further improvements in the precision experiments
in order to identify a signal of new physics beyond the SM.
We find that the two polarization asymmetries at high energies,
$P_\tau$ and $A_{\rm LR}$, are most effective in this regard since
they constrain the parameter $\sbar^2(m_Z^2)$ directly without
suffering from the QCD uncertainty.
At low energies, two polarization experiments in the $e$--$q$ sector,
the polarized $e$D scattering and the APV measurements, may have
the potential of identifying physics beyond the
${\rm SU(2)_L \times U(1)_Y}$ universality.
We should note, however, that a better measurement of the hadronic
vacuum polarization effect
$\delta_\alpha =\delta[(\Delta\frac{1}{\alpha})_{\rm hadrons}]$ is needed
in order for us to look beyond the SM through the electroweak
radiative effects.

\section*{\small {\bf ACKNOWLEDGEMENTS }}
\pr
We would like to thank M.~Drees, D.~Haidt, F.~Halzen, B.~Kniehl, P.~Langacker,
N.A.~McDougall, H.~Tanabashi, Y.~Yamada and D.~Zeppenfeld for discussions.
We learned a lot from T.~Mori and H.~Masuda on the experiments being
performed at LEP and SLC.
\baselineskip10pt

\def\pl #1 #2 #3 {Phys.~Lett. {\bf#1}, #2 (#3)}
\def\np #1 #2 #3 {Nucl.~Phys. {\bf#1}, #2 (#3)}
\def\zp #1 #2 #3 {Z.~Phys. {\bf#1}, #2 (#3)}
\def\pr #1 #2 #3 {Phys.~Rev. {\bf#1}, #2 (#3)}
\def\prep #1 #2 #3 {Phys.~Rep. {\bf#1}, #2 (#3)}
\def\prl #1 #2 #3 {Phys.~Rev.~Lett. {\bf#1}, #2 (#3)}
\def\mpl #1 #2 #3 {Mod.~Phys.~Lett. {\bf#1}, #2 (#3)}
\def\ptp #1 #2 #3 {Prog.~Theor.~Phys. {\bf#1}, #2 (#3)}
\def\xx #1 #2 #3 {{\bf#1}, #2 (#3)}
\def\etal {{\it et al}.}
\def\eg {{\it e.g}.}
\def\ie {{\it i.e}.}
\def\ibid{{\it ibid}. }

\end{document}